\documentclass[onecolumn,letterpaper,aps,prc,longbibliography,superscriptaddress,nofootinbib,floatfix]{revtex4-2}
\usepackage{graphicx}
\usepackage{lineno}
\begin{document}


\title{Highlights from the PHENIX experiment (Quark Matter 2022) \\
Sanghoon Lim, Pusan National University, \\
(for the PHENIX Collaboration)}

%\thanks{Presented at Quark Matter 2022}%

\newcommand{\abilene}{Abilene Christian University, Abilene, Texas 79699, USA}
\newcommand{\augie}{Department of Physics, Augustana University, Sioux Falls, South Dakota 57197, USA}
\newcommand{\banaras}{Department of Physics, Banaras Hindu University, Varanasi 221005, India}
\newcommand{\barc}{Bhabha Atomic Research Centre, Bombay 400 085, India}
\newcommand{\baruch}{Baruch College, City University of New York, New York, New York, 10010 USA}
\newcommand{\bnlcoll}{Collider-Accelerator Department, Brookhaven National Laboratory, Upton, New York 11973-5000, USA}
\newcommand{\bnlphys}{Physics Department, Brookhaven National Laboratory, Upton, New York 11973-5000, USA}
\newcommand{\caucr}{University of California-Riverside, Riverside, California 92521, USA}
\newcommand{\charlesczech}{Charles University, Faculty of Mathematics and Physics, 180 00 Troja, Prague, Czech Republic}
\newcommand{\ciae}{Science and Technology on Nuclear Data Laboratory, China Institute of Atomic Energy, Beijing 102413, People's Republic of China}
\newcommand{\cns}{Center for Nuclear Study, Graduate School of Science, University of Tokyo, 7-3-1 Hongo, Bunkyo, Tokyo 113-0033, Japan}
\newcommand{\colorado}{University of Colorado, Boulder, Colorado 80309, USA}
\newcommand{\columbia}{Columbia University, New York, New York 10027 and Nevis Laboratories, Irvington, New York 10533, USA}
\newcommand{\czechtech}{Czech Technical University, Zikova 4, 166 36 Prague 6, Czech Republic}
\newcommand{\dapnia}{Dapnia, CEA Saclay, F-91191, Gif-sur-Yvette, France}
\newcommand{\debrecen}{Debrecen University, H-4010 Debrecen, Egyetem t{\'e}r 1, Hungary}
\newcommand{\elte}{ELTE, E{\"o}tv{\"o}s Lor{\'a}nd University, H-1117 Budapest, P{\'a}zm{\'a}ny P.~s.~1/A, Hungary}
\newcommand{\ewha}{Ewha Womans University, Seoul 120-750, Korea}
\newcommand{\famu}{Florida A\&M University, Tallahassee, FL 32307, USA}
\newcommand{\fsu}{Florida State University, Tallahassee, Florida 32306, USA}
\newcommand{\gsu}{Georgia State University, Atlanta, Georgia 30303, USA}
\newcommand{\hanyang}{Hanyang University, Seoul 133-792, Korea}
\newcommand{\hiroshima}{Hiroshima University, Kagamiyama, Higashi-Hiroshima 739-8526, Japan}
\newcommand{\howard}{Department of Physics and Astronomy, Howard University, Washington, DC 20059, USA}
\newcommand{\ihepprot}{IHEP Protvino, State Research Center of Russian Federation, Institute for High Energy Physics, Protvino, 142281, Russia}
\newcommand{\illuiuc}{University of Illinois at Urbana-Champaign, Urbana, Illinois 61801, USA}
\newcommand{\inrras}{Institute for Nuclear Research of the Russian Academy of Sciences, prospekt 60-letiya Oktyabrya 7a, Moscow 117312, Russia}
\newcommand{\instpasczech}{Institute of Physics, Academy of Sciences of the Czech Republic, Na Slovance 2, 182 21 Prague 8, Czech Republic}
\newcommand{\isu}{Iowa State University, Ames, Iowa 50011, USA}
\newcommand{\jaea}{Advanced Science Research Center, Japan Atomic Energy Agency, 2-4 Shirakata Shirane, Tokai-mura, Naka-gun, Ibaraki-ken 319-1195, Japan}
\newcommand{\jeonbuk}{Jeonbuk National University, Jeonju, 54896, Korea}
\newcommand{\jyvaskyla}{Helsinki Institute of Physics and University of Jyv{\"a}skyl{\"a}, P.O.Box 35, FI-40014 Jyv{\"a}skyl{\"a}, Finland}
\newcommand{\kek}{KEK, High Energy Accelerator Research Organization, Tsukuba, Ibaraki 305-0801, Japan}
\newcommand{\korea}{Korea University, Seoul 02841, Korea}
\newcommand{\kurchatov}{National Research Center ``Kurchatov Institute", Moscow, 123098 Russia}
\newcommand{\kyoto}{Kyoto University, Kyoto 606-8502, Japan}
\newcommand{\labllr}{Laboratoire Leprince-Ringuet, Ecole Polytechnique, CNRS-IN2P3, Route de Saclay, F-91128, Palaiseau, France}
\newcommand{\lahorelums}{Physics Department, Lahore University of Management Sciences, Lahore 54792, Pakistan}
\newcommand{\lawllnl}{Lawrence Livermore National Laboratory, Livermore, California 94550, USA}
\newcommand{\losalamos}{Los Alamos National Laboratory, Los Alamos, New Mexico 87545, USA}
\newcommand{\lpc}{LPC, Universit{\'e} Blaise Pascal, CNRS-IN2P3, Clermont-Fd, 63177 Aubiere Cedex, France}
\newcommand{\lund}{Department of Physics, Lund University, Box 118, SE-221 00 Lund, Sweden}
\newcommand{\lyon}{IPNL, CNRS/IN2P3, Univ Lyon, Université Lyon 1, F-69622, Villeurbanne, France}
\newcommand{\maryland}{University of Maryland, College Park, Maryland 20742, USA}
\newcommand{\mass}{Department of Physics, University of Massachusetts, Amherst, Massachusetts 01003-9337, USA}
\newcommand{\mate}{MATE, Laboratory of Femtoscopy, K\'aroly R\'obert Campus, Gy\"ongy\"os, Hungary}
\newcommand{\michigan}{Department of Physics, University of Michigan, Ann Arbor, Michigan 48109-1040, USA}
\newcommand{\miss}{Mississippi State University, Mississippi State, Mississippi 39762, USA}
\newcommand{\muenster}{Institut f\"ur Kernphysik, University of M\"unster, D-48149 M\"unster, Germany}
\newcommand{\muhlenberg}{Muhlenberg College, Allentown, Pennsylvania 18104-5586, USA}
\newcommand{\myongji}{Myongji University, Yongin, Kyonggido 449-728, Korea}
\newcommand{\nagasaki}{Nagasaki Institute of Applied Science, Nagasaki-shi, Nagasaki 851-0193, Japan}
\newcommand{\nara}{Nara Women's University, Kita-uoya Nishi-machi Nara 630-8506, Japan}
\newcommand{\natmephi}{National Research Nuclear University, MEPhI, Moscow Engineering Physics Institute, Moscow, 115409, Russia}
\newcommand{\newmex}{University of New Mexico, Albuquerque, New Mexico 87131, USA}
\newcommand{\nmsu}{New Mexico State University, Las Cruces, New Mexico 88003, USA}
\newcommand{\northcg}{Physics and Astronomy Department, University of North Carolina at Greensboro, Greensboro, North Carolina 27412, USA}
\newcommand{\ohio}{Department of Physics and Astronomy, Ohio University, Athens, Ohio 45701, USA}
\newcommand{\ornl}{Oak Ridge National Laboratory, Oak Ridge, Tennessee 37831, USA}
\newcommand{\orsay}{IPN-Orsay, Univ.~Paris-Sud, CNRS/IN2P3, Universit\'e Paris-Saclay, BP1, F-91406, Orsay, France}
\newcommand{\peking}{Peking University, Beijing 100871, People's Republic of China}
\newcommand{\pnpi}{PNPI, Petersburg Nuclear Physics Institute, Gatchina, Leningrad region, 188300, Russia}
\newcommand{\pusan}{Pusan National University, Pusan 46241, Korea}
\newcommand{\riken}{RIKEN Nishina Center for Accelerator-Based Science, Wako, Saitama 351-0198, Japan}
\newcommand{\rikjrbrc}{RIKEN BNL Research Center, Brookhaven National Laboratory, Upton, New York 11973-5000, USA}
\newcommand{\rikkyo}{Physics Department, Rikkyo University, 3-34-1 Nishi-Ikebukuro, Toshima, Tokyo 171-8501, Japan}
\newcommand{\saispbstu}{Saint Petersburg State Polytechnic University, St.~Petersburg, 195251 Russia}
\newcommand{\saopaulo}{Universidade de S{\~a}o Paulo, Instituto de F\'{\i}sica, Caixa Postal 66318, S{\~a}o Paulo CEP05315-970, Brazil}
\newcommand{\seoulnat}{Department of Physics and Astronomy, Seoul National University, Seoul 151-742, Korea}
\newcommand{\stonybrkc}{Chemistry Department, Stony Brook University, SUNY, Stony Brook, New York 11794-3400, USA}
\newcommand{\stonycrkp}{Department of Physics and Astronomy, Stony Brook University, SUNY, Stony Brook, New York 11794-3800, USA}
\newcommand{\sungskku}{Sungkyunkwan University, Suwon, 440-746, Korea}
\newcommand{\tenn}{University of Tennessee, Knoxville, Tennessee 37996, USA}
\newcommand{\texsu}{Texas Southern University, Houston, TX 77004, USA}
\newcommand{\titech}{Department of Physics, Tokyo Institute of Technology, Oh-okayama, Meguro, Tokyo 152-8551, Japan}
\newcommand{\tsukuba}{Tomonaga Center for the History of the Universe, University of Tsukuba, Tsukuba, Ibaraki 305, Japan}
\newcommand{\vandy}{Vanderbilt University, Nashville, Tennessee 37235, USA}
\newcommand{\weizmann}{Weizmann Institute, Rehovot 76100, Israel}
\newcommand{\wigner}{Institute for Particle and Nuclear Physics, Wigner Research Centre for Physics, Hungarian Academy of Sciences (Wigner RCP, RMKI) H-1525 Budapest 114, POBox 49, Budapest, Hungary}
\newcommand{\yonsei}{Yonsei University, IPAP, Seoul 120-749, Korea}
\newcommand{\zagreb}{Department of Physics, Faculty of Science, University of Zagreb, Bijeni\v{c}ka c.~32 HR-10002 Zagreb, Croatia}
\newcommand{\zambia}{Department of Physics, School of Natural Sciences, University of Zambia, Great East Road Campus, Box 32379, Lusaka, Zambia}
\affiliation{\abilene}
\affiliation{\augie}
\affiliation{\banaras}
\affiliation{\barc}
\affiliation{\baruch}
\affiliation{\bnlcoll}
\affiliation{\bnlphys}
\affiliation{\caucr}
\affiliation{\charlesczech}
\affiliation{\ciae}
\affiliation{\cns}
\affiliation{\colorado}
\affiliation{\columbia}
\affiliation{\czechtech}
\affiliation{\dapnia}
\affiliation{\debrecen}
\affiliation{\elte}
\affiliation{\ewha}
\affiliation{\famu}
\affiliation{\fsu}
\affiliation{\gsu}
\affiliation{\hanyang}
\affiliation{\hiroshima}
\affiliation{\howard}
\affiliation{\ihepprot}
\affiliation{\illuiuc}
\affiliation{\inrras}
\affiliation{\instpasczech}
\affiliation{\isu}
\affiliation{\jaea}
\affiliation{\jeonbuk}
\affiliation{\jyvaskyla}
\affiliation{\kek}
\affiliation{\korea}
\affiliation{\kurchatov}
\affiliation{\kyoto}
\affiliation{\labllr}
\affiliation{\lahorelums}
\affiliation{\lawllnl}
\affiliation{\losalamos}
\affiliation{\lpc}
\affiliation{\lund}
\affiliation{\lyon}
\affiliation{\maryland}
\affiliation{\mass}
\affiliation{\mate}
\affiliation{\michigan}
\affiliation{\miss}
\affiliation{\muenster}
\affiliation{\muhlenberg}
\affiliation{\myongji}
\affiliation{\nagasaki}
\affiliation{\nara}
\affiliation{\natmephi}
\affiliation{\newmex}
\affiliation{\nmsu}
\affiliation{\northcg}
\affiliation{\ohio}
\affiliation{\ornl}
\affiliation{\orsay}
\affiliation{\peking}
\affiliation{\pnpi}
\affiliation{\pusan}
\affiliation{\riken}
\affiliation{\rikjrbrc}
\affiliation{\rikkyo}
\affiliation{\saispbstu}
\affiliation{\saopaulo}
\affiliation{\seoulnat}
\affiliation{\stonybrkc}
\affiliation{\stonycrkp}
\affiliation{\sungskku}
\affiliation{\tenn}
\affiliation{\texsu}
\affiliation{\titech}
\affiliation{\tsukuba}
\affiliation{\vandy}
\affiliation{\weizmann}
\affiliation{\wigner}
\affiliation{\yonsei}
\affiliation{\zagreb}
\affiliation{\zambia}
\author{N.J.~Abdulameer} \affiliation{\debrecen}
\author{U.~Acharya} \affiliation{\gsu} 
\author{A.~Adare} \affiliation{\colorado} 
\author{C.~Aidala} \affiliation{\losalamos} \affiliation{\michigan} 
\author{N.N.~Ajitanand} \altaffiliation{Deceased} \affiliation{\stonybrkc} 
\author{Y.~Akiba} \email[PHENIX Spokesperson: ]{akiba@rcf.rhic.bnl.gov} \affiliation{\riken} \affiliation{\rikjrbrc} 
\author{R.~Akimoto} \affiliation{\cns} 
\author{H.~Al-Ta'ani} \affiliation{\nmsu} 
\author{J.~Alexander} \affiliation{\stonybrkc} 
\author{M.~Alfred} \affiliation{\howard} 
% ...
\collaboration{PHENIX Collaboration}  \noaffiliation

%\linenumbers

\begin{abstract}

PHENIX has performed an extensive study on the evolution of medium effects 
from small to large systems. PHENIX has continued searching for Quark-Gluon 
Plasma (QGP) in small systems by measuring collectivity, modification of 
light hadron and quarkonia production, and jet substructure. In large 
systems, detailed studies on the property of the QGP have been done using 
direct photon, $\pi^{0}$-hadron correlation, heavy-flavor electron, and 
$J/\psi$ flow with a large statistics of data collected in 2014. This report 
covers new results from the PHENIX experiment in various collision systems.

\end{abstract}

\maketitle

\section{Searching for QGP in small systems}

\subsection{Collectivity}

\begin{figure}[htb]
\centerline{%
\includegraphics[width=0.9\textwidth]{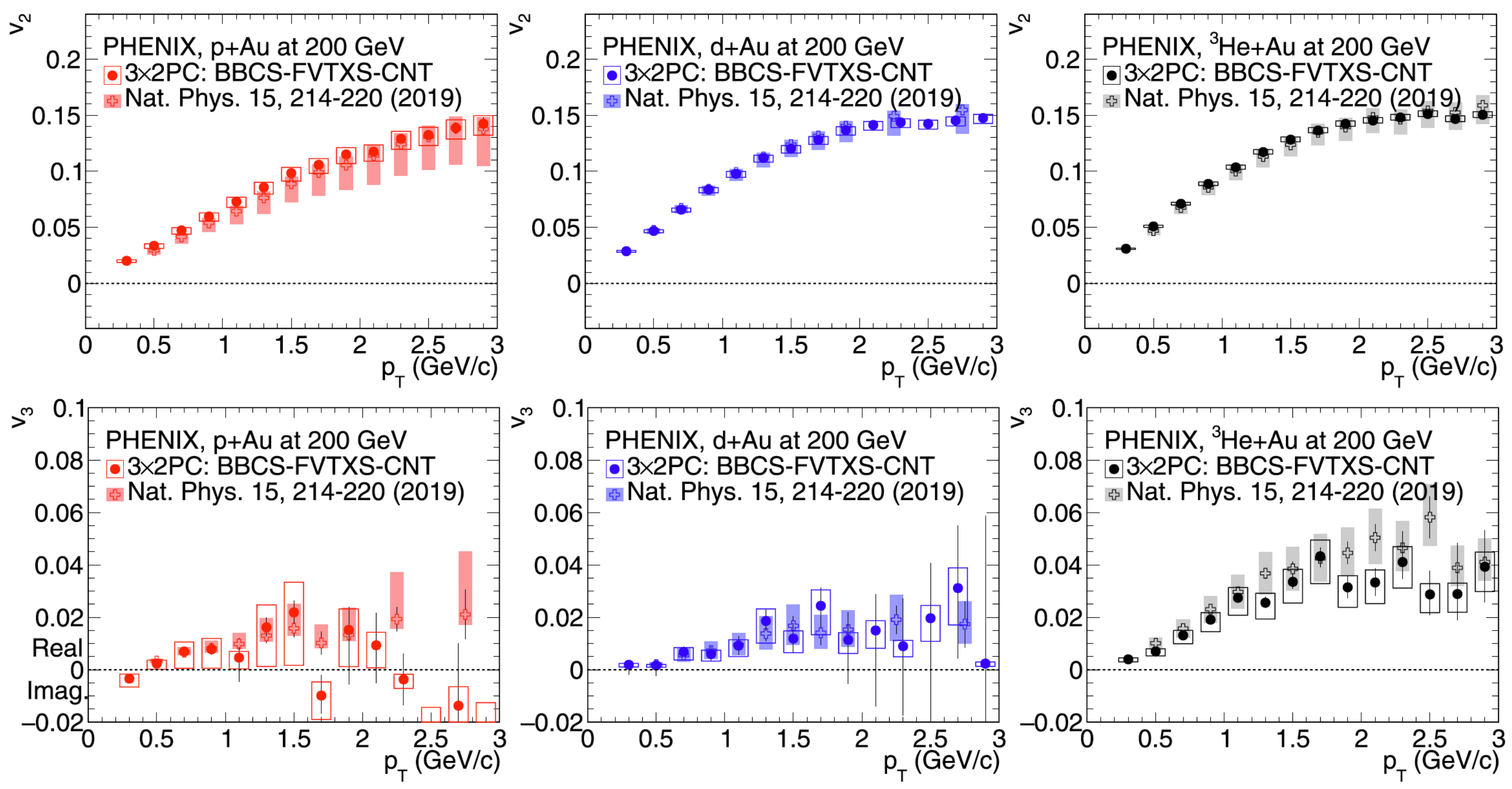}}
\caption{Elliptic (top) and triangular flow of charged particles as a function of $p_T$ at mid-rapidity in 0--5\% $p$/$d$/$^{3}$He+Au collisions at $\sqrt{s_{NN}}=200$ GeV.}
\label{Fig:small_flow}
\end{figure}

PHENIX previously published results of elliptic and triangular flow in 
0--5\% $p$/$d$/$^{3}$He+Au collisions at $\sqrt{s_{NN}}=200$ 
GeV~\cite{PHENIX:2018lia}. The results show a strong relation between 
initial geometry and flow which is evidence of QGP droplets in small 
systems. A new analysis has been performed using three combinations of 
two-particle correlations~\cite{PHENIX:2021ubk}. Figure~\ref{Fig:small_flow} 
shows the elliptic and triangular flow of charged particles as a function of 
$p_T$ in 0--5\% $p$/$d$/$^{3}$He+Au collisions using two different methods. 
The new results with the $3\times2$PC method are consistent with the 
previous results with the event-plane method. More detailed studies are done 
to extract flow coefficients at various kinematic regions, and such a study 
can help to understand the origin of azimuthal anisotropies like flow, 
nonflow, and longitudinal decorrelation in different kinematic 
regions~\cite{PHENIX:2021ubk,PHENIX:2022nht}.

\subsection{Nuclear modification of light-flavor hadrons}

\begin{figure}[htb]
\centerline{%
\includegraphics[width=0.9\textwidth]{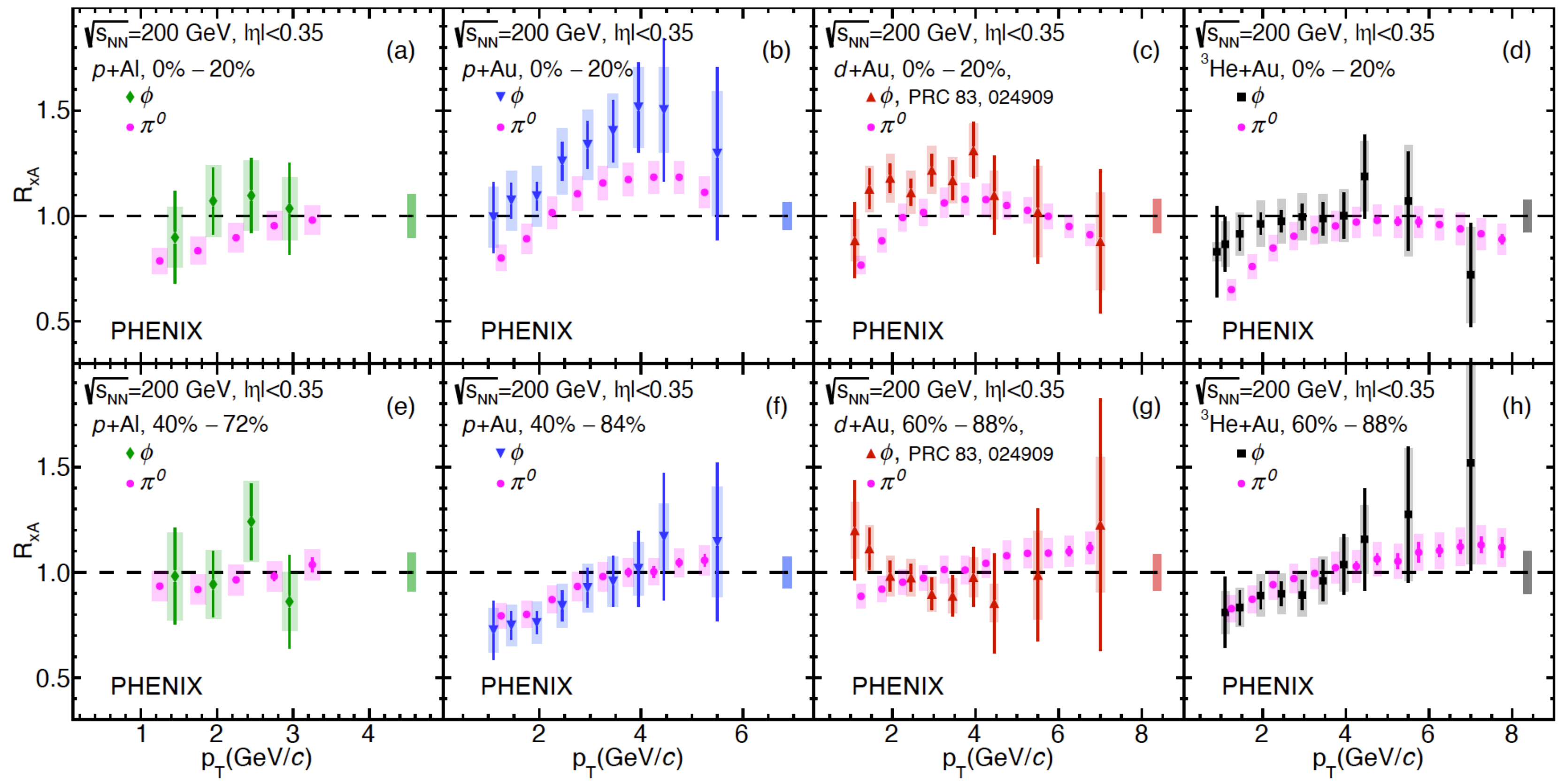}}
\caption{Nuclear modification factor of $\phi$ and $\pi^{0}$ as a function of $p_T$ in central (top) and peripheral (bottom) collisions of small systems at $\sqrt{s_{NN}}=200$ GeV.}
\label{Fig:small_phi}
\end{figure}

A study of production of light-flavor hadrons in various centrality ranges of small collision systems can provide detailed information on the evolution of initial-state and final-state effects.
Figure~\ref{Fig:small_phi} shows the nuclear modification factor of $\pi^{0}$ and $\phi$ as a function of $p_T$ in central and peripheral collisions~\cite{PHENIX:2021dod,PHENIX:2022hku}.
In central collisions, there is a $p_T$ broadening in $2<p_T<6~\mathrm{GeV}/c$, and it is more pronounced in $p$+Au collisions compared to $p$+Al collisions of a different target and $^{3}$He+Au collisions of a different projectile.
The radial flow could explain this $p_T$ broadening.
In the comparison between $\pi^{0}$ and $\phi$, the nuclear modification factor of $\phi$ is slightly above than $\pi^{0}$ in central collisions.
This indicates a multiplicity-dependent strangeness enhancement in small systems.

\begin{figure}[htb]
\centerline{%
\includegraphics[width=0.9\textwidth]{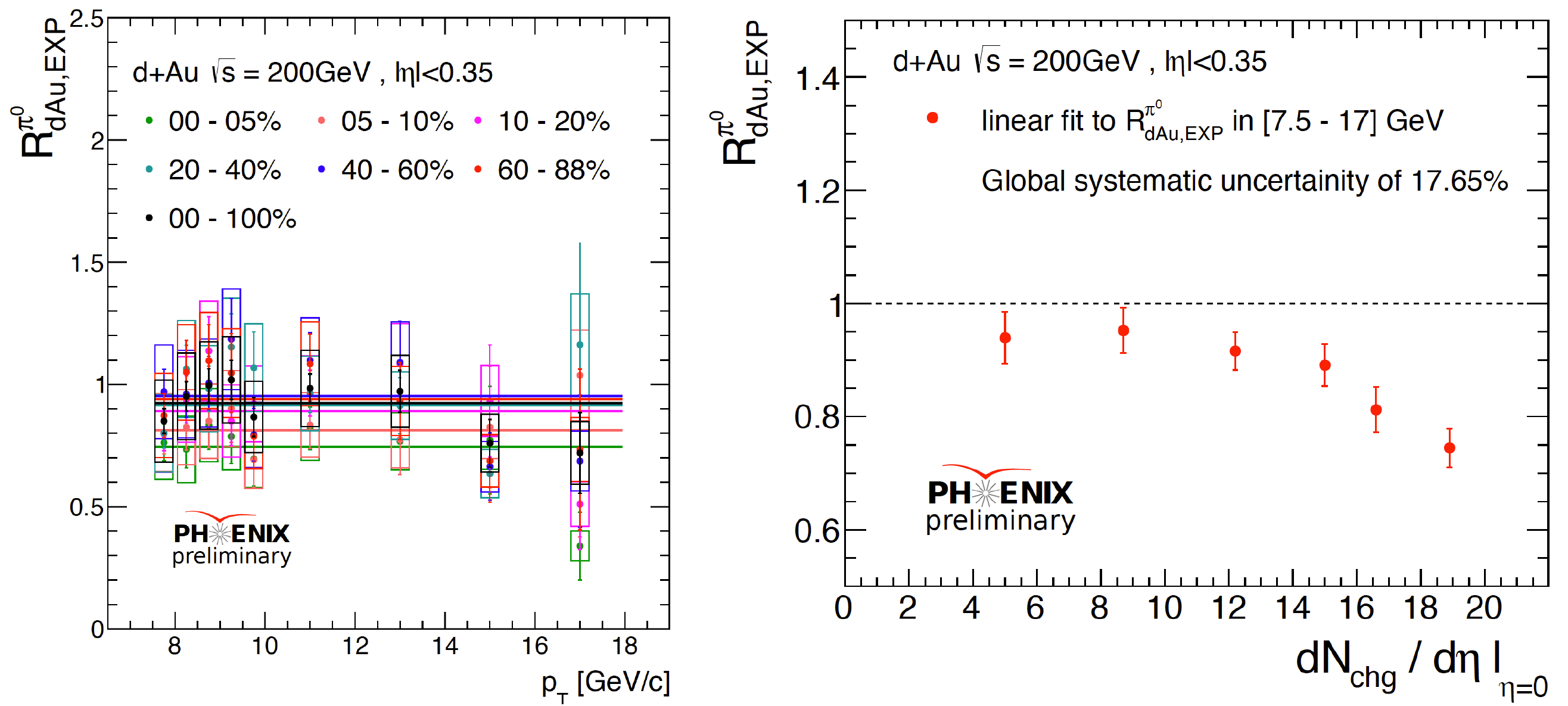}}
\caption{Nuclear modification factor of $\pi^{0}$ with respect to the nuclear modification of direct $\gamma$ at high $p_T$ in various centrality of $d$+Au collisions at $\sqrt{s_{NN}}=200$ GeV.}
\label{Fig:small_highpt}
\end{figure}

One interesting observation in the nuclear modification factor of $\pi^{0}$ at high $p_T$, there is a suppression (enhancement) in central (peripheral) collisions.
To investigate the origin of the modification, PHENIX has measured the nuclear modification factor of $\pi^{0}$ and direct $\gamma$ at high $p_T$ in various centrality of $d$+Au collisions at $\sqrt{s_{NN}}=200$ GeV.
A similar enhancement is observed in the nuclear modification factor of direct $\gamma$ in peripheral collisions, but there is no suppression in central collisions.
To eliminate a possible bias from centrality selection and $\left< N_\mathrm{coll} \right>$ calculation, the ratio of nuclear modification factors between $\pi^{0}$ and direct $\gamma$ is calculated as shown in the left panel of Fig.~\ref{Fig:small_highpt}.
The right panel shows the quantity as a function of charged particle multiplicity.
In high multiplicity $d$+Au collisions, there is a clear indication of nuclear effects suppressing $\pi^{0}$ production.

\subsection{Charmonia production}

\begin{figure}[htb]
\centerline{%
\includegraphics[width=0.9\textwidth]{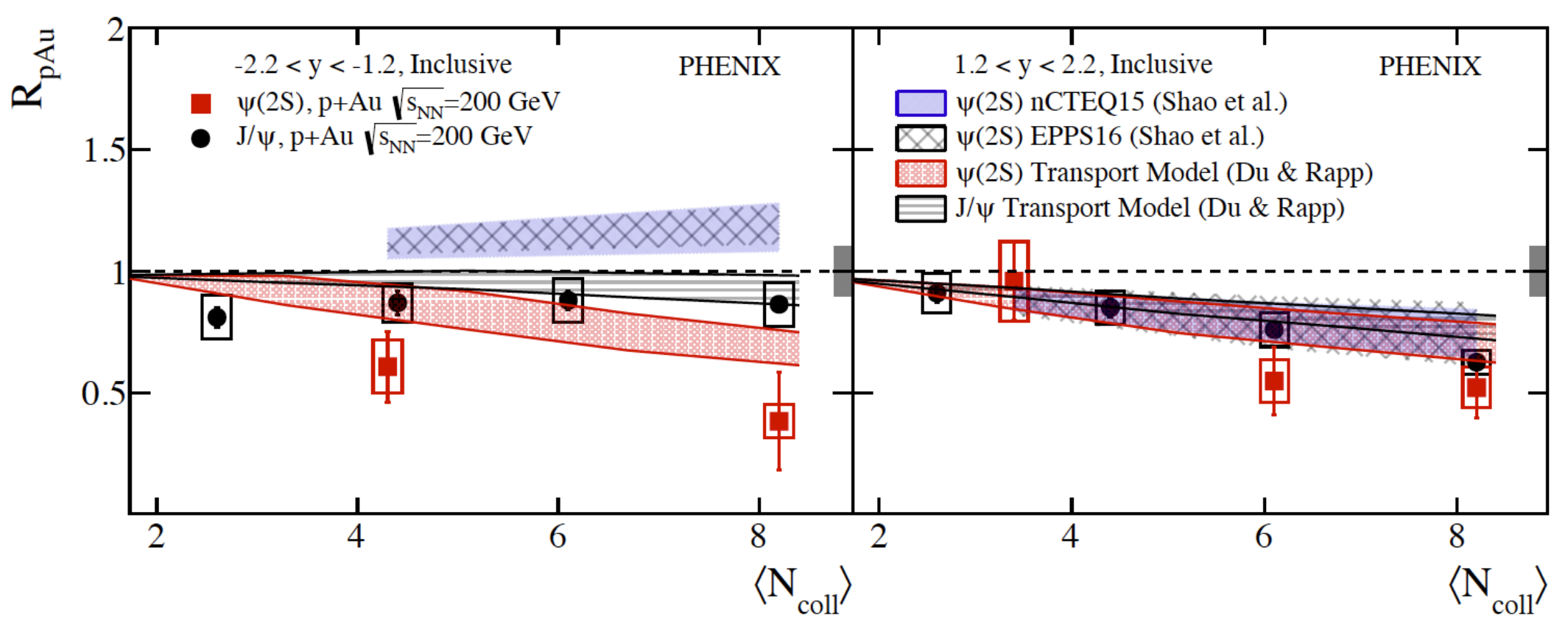}}
\caption{Nuclear modification of $J/\psi$ and $\psi(2S)$ at backward (left, Au-direction) and forward (right, $p$-direction) rapidity as a function of $\left< N_\mathrm{coll} \right>$ in $p$+Au collisions at $\sqrt{s_{NN}}=200$ GeV.}
\label{Fig:small_psi}
\end{figure}

PHENIX previously published the nuclear modification of inclusive $J/\psi$ in various small collision systems~\cite{PHENIX:2019brm}.
The suppression of $J/\psi$ at backward rapidity (Au-direction), which nuclear parton distribution functions (nPDFs) could not explain, suggests final-state effects on the $J/\psi$ production. 
A new analysis has been done to measure the nuclear modification factor of $\psi(2S)$, which could provide additional information on final-state effects compared with the $J/\psi$ results.
This is because it is expected that initial-state effects on $J/\psi$ and $\psi(2S)$ production are similar.
Figure~\ref{Fig:small_psi} shows the nuclear modification factor of $J/\psi$ and $\psi(2S)$ at backward (left) and forward (right) rapidity as a function of centrality in $p$+Au collisions at $\sqrt{s_{NN}}=200$ GeV~\cite{PHENIX:2022nrm}.
The modification of $J/\psi$ and $\psi(2S)$ is similar at forward rapidity, which can be explained by models including nPDFs.
At backward rapidity, $\psi(2S)$ is more strongly suppressed than $J/\psi$ in central collisions.
The transport model considering final-state effects can describe the data qualitatively.

\section{Detailed study on QGP in large systems}

\subsection{Direct photon}

\begin{figure}[htb]
\centerline{%
\includegraphics[width=0.9\textwidth]{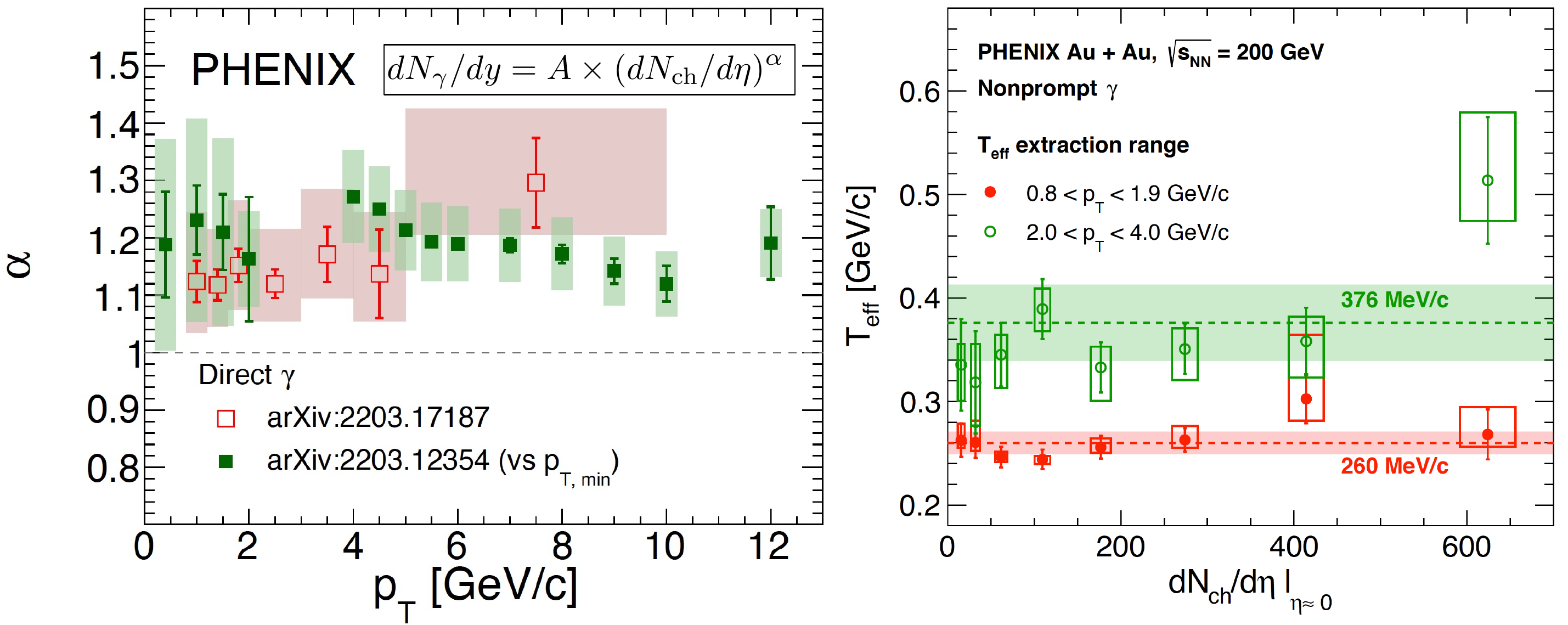}}
\caption{Scaling factors $\alpha$ as a function of direct photon $p_T$ and inverse slope parameters from non-prompt photon $p_T$ spectra as a function of charged particle multiplicity in two $p_T$ ranges.}
\label{Fig:large_directg}
\end{figure}

Direct photons are important probes to understand the evolution of the medium produced in heavy-ion collisions.
PHENIX previously showed that the yields of direct photons at low $p_T$ scale with the charged particle multiplicity ($dN_{\gamma}/dy = A \times (dN_\mathrm{ch}/d\eta)^{\alpha}$), and the scaling factor $\alpha$ is about 1.25 for yields in heavy-ion collisions at $\sqrt{s_{NN}}=39$--200 GeV.
The scaling factors in small systems, $p$+Au and $d$+Au collisions, are located in between $p$+$p$ and heavy-ion results.
A new analysis has been performed with enhanced statistics of Au+Au data~\cite{PHENIX:2022qfp,PHENIX:2022rsx}.
The left panel of Fig.~\ref{Fig:large_directg} shows the scaling factor $\alpha$ as a function of direct photon $p_T$ in the integrated centrality of Au+Au collisions.
Interestingly, the scaling factor weakly depends on $p_T$, although the source of direct photons varies with $p_T$.
When comparing with direct photon yields at 2.76 TeV~\cite{ALICE:2015xmh}, a good scaling with multiplicity is observed for $p_T<2~\mathrm{GeV}/c$ from 39 GeV to 2.76 TeV.
At $p_T>2~\mathrm{GeV}/c$, the yield at 2.76 TeV is larger than the scaled yield based on the RHIC energies.
This indicates that the medium temperature is higher at the LHC energy.

For more investigation on the sources of direct photons at low $p_T$, the non-prompt direct photon yields are calculated by subtracting the scaled $p$+$p$ yield~\cite{PHENIX:2022rsx}.
The $p_T$ spectra of non-prompt direct photons are not described by a single exponential function ($\exp(-p_T/T_\mathrm{eff})$), so the inverse slope parameters are extracted at two $p_T$ ranges, $0.8<p_T<1.9~\mathrm{GeV}/c$ and $2<p_T<4~\mathrm{GeV}/c$, for each non-prompt direct photon $p_T$ spectrum.
The right panel of Fig.~\ref{Fig:large_directg} shows the inverse slope parameter as a function of multiplicity.
A higher value is obtained at the higher $p_T$ range, indicating non-prompt direct photons in different $p_T$ are produced at different stages, and there is no significant multiplicity dependence.

\subsection{$\pi^{0}$--$h$ correlation}

\begin{figure}[htb]
\centerline{%
\includegraphics[width=0.9\textwidth]{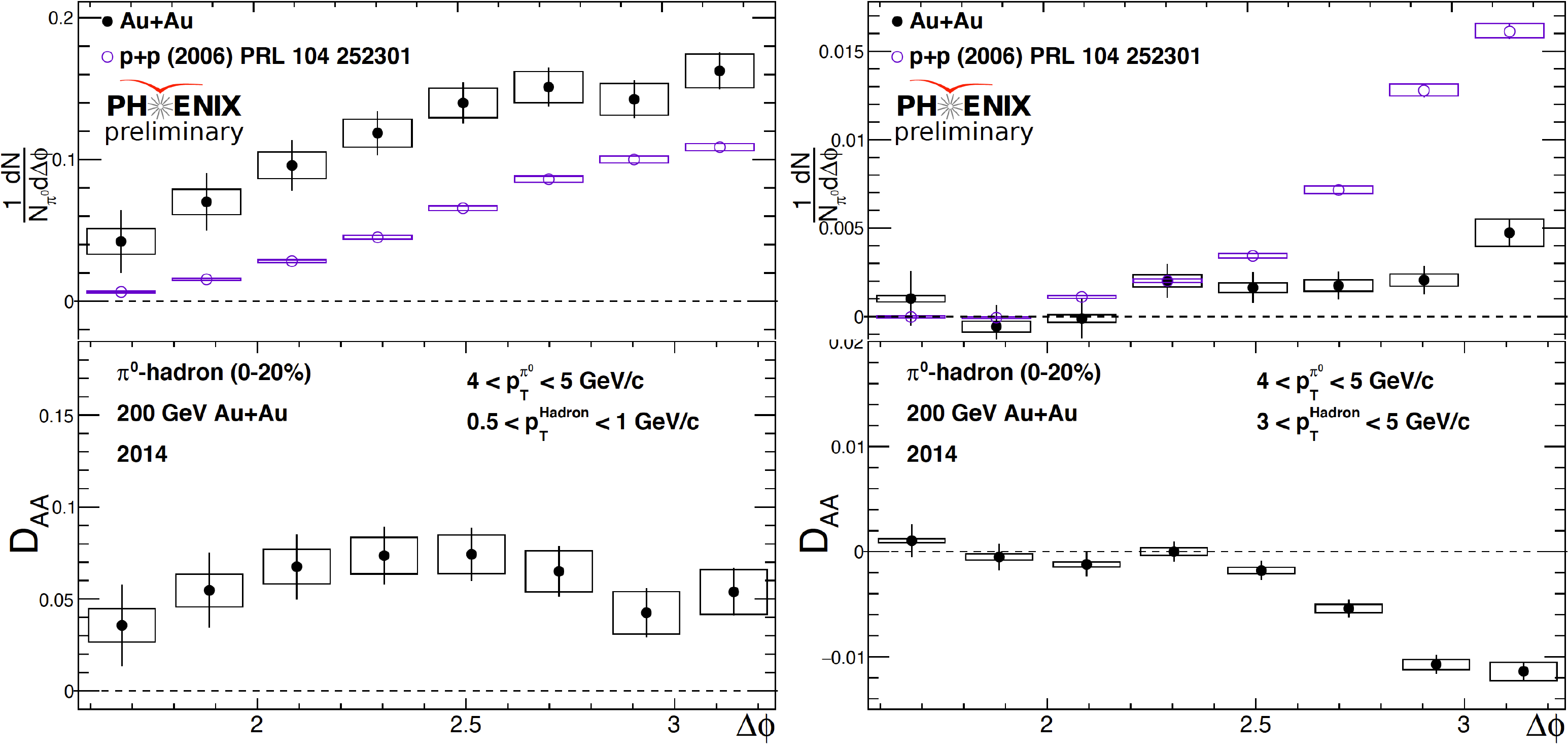}}
\caption{Per trigger pair yields of $\pi^{0}$-$h$ correlation in $p$+$p$ and central Au+Au collisions at $\sqrt{s_{NN}}=200$ GeV for two $p_T$ ranges of associated particles and the difference between results in two systems.}
\label{Fig:large_daa}
\end{figure}

The modification of jet shape inside the QGP can be studied via the azimuthal correlation between trigger particles at high $p_T$ and associated particles in various $p_T$ ranges.
New preliminary results on $\pi^{0}$--$h$ correlations have been obtained with high statistics data collected in 2014.
Figure~\ref{Fig:large_daa} shows per trigger yields as a function of $\Delta\phi$ of $\pi^{0}$-$h$ correlation in $p$+$p$ and central Au+Au collisions.
Trigger particles are in $4<p_T<5~\mathrm{GeV}/c$, and associated particles are in two $p_T$ ranges, $0.5<p_T<1~\mathrm{GeV}/c$ (left) and $3<p_T<5~\mathrm{GeV}/c$ (right), are shown.
The bottom panels show the difference in the per-trigger yields between the two systems.
In the results for low $p_T$ associated particles, the yields in Au+Au are larger than those in $p$+$p$ at the broad $\Delta\phi$ range.
However, the yield in Au+Au is smaller at $\Delta\phi\sim\pi$ for associated particles in the higher $p_T$ region.
The results suggest a picture that high $p_T$ partons around the core of jets lose energy, and the energy is redistributed broadly to low $p_T$ partons.

\subsection{Electrons from charm and bottom decay}

\begin{figure}[htb]
\centerline{%
\includegraphics[width=0.85\textwidth]{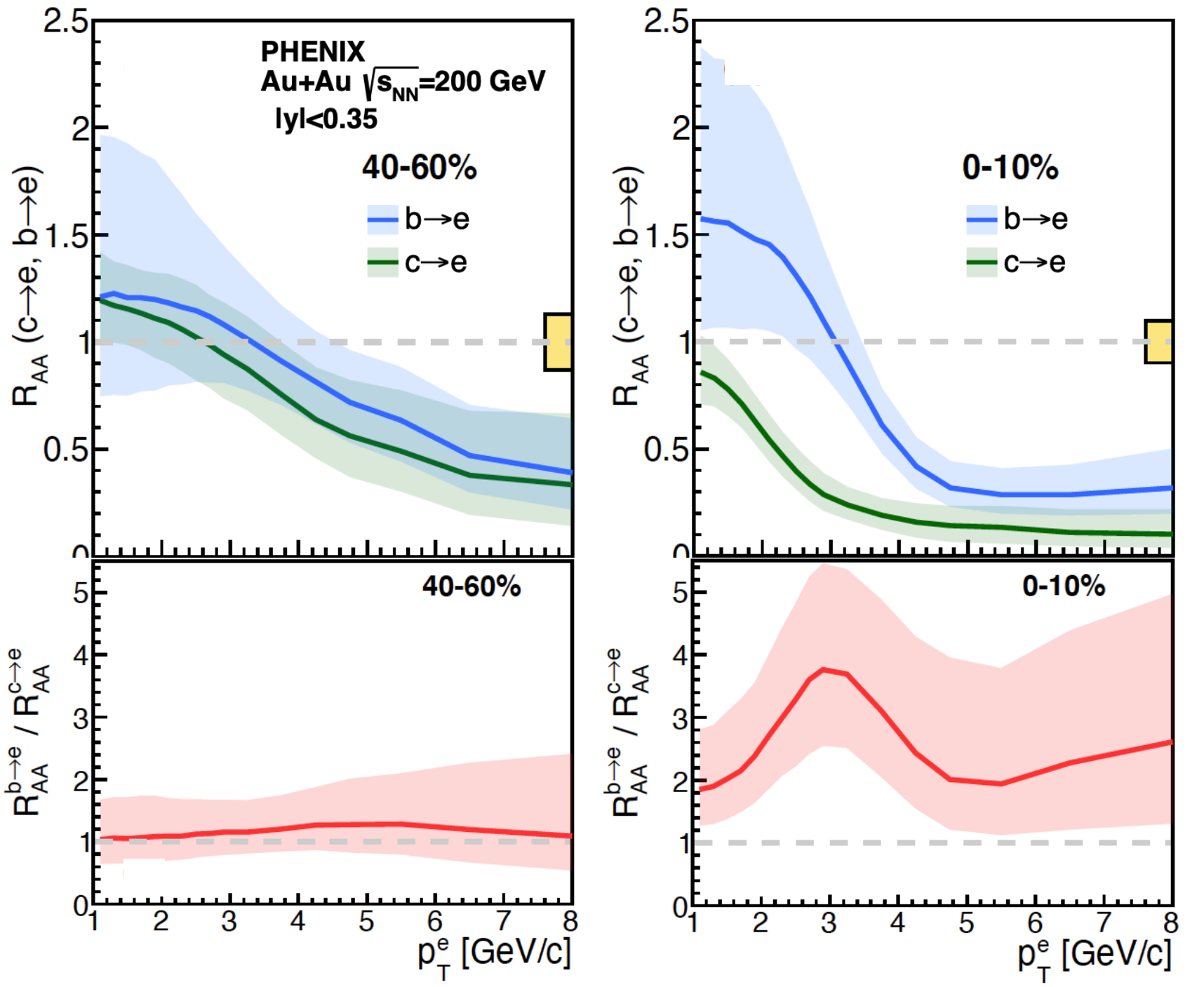}}
\caption{Nuclear modification factor of electrons from the charm and bottom hadron decays in peripheral (left) and central (right) Au+Au collisions at $\sqrt{s_{NN}}=200$ GeV. The ratio of electron $R_{AA}$ from bottom hadron to charm hadron is presented in the bottom panels.}
\label{Fig:large_hfe}
\end{figure}

Heavy quarks are useful tools to investigate the medium properties of the QGP. 
The modification of charm and bottom quarks are expected to be different, and one possible reason is the dead-cone effect.
PHENIX has measured yields of electrons from the charm and bottom hadron decays both in $p$+$p$ and Au+Au collisions at $\sqrt{s_{NN}}=200$ GeV~\cite{PHENIX:2019pxh,PHENIX:2022wim}, and the nuclear modification factor ($R_{AA}$) is calculated with the yields obtained using the same analysis method.
Figure~\ref{Fig:large_hfe} shows $R_{AA}$ of electrons from charm (green curves) and bottom (blue curves) hadron decays in peripheral (left) and central (right) Au+Au collisions $\sqrt{s_{NN}}=200$ GeV, and the ratio of bottom electron $R_{AA}$ to charm electron $R_{AA}$ is presented in the bottom panels.
In peripheral collisions, the nuclear modification factor for charm and bottom electrons is consistent within the uncertainties.
However, the modification is more substantial for charm electrons than bottom electrons in central collisions indicating quark mass-dependent energy loss inside the QGP.

\subsection{$J/\psi$ elliptic flow}

\begin{figure}[htb]
\centerline{%
\includegraphics[width=0.45\textwidth]{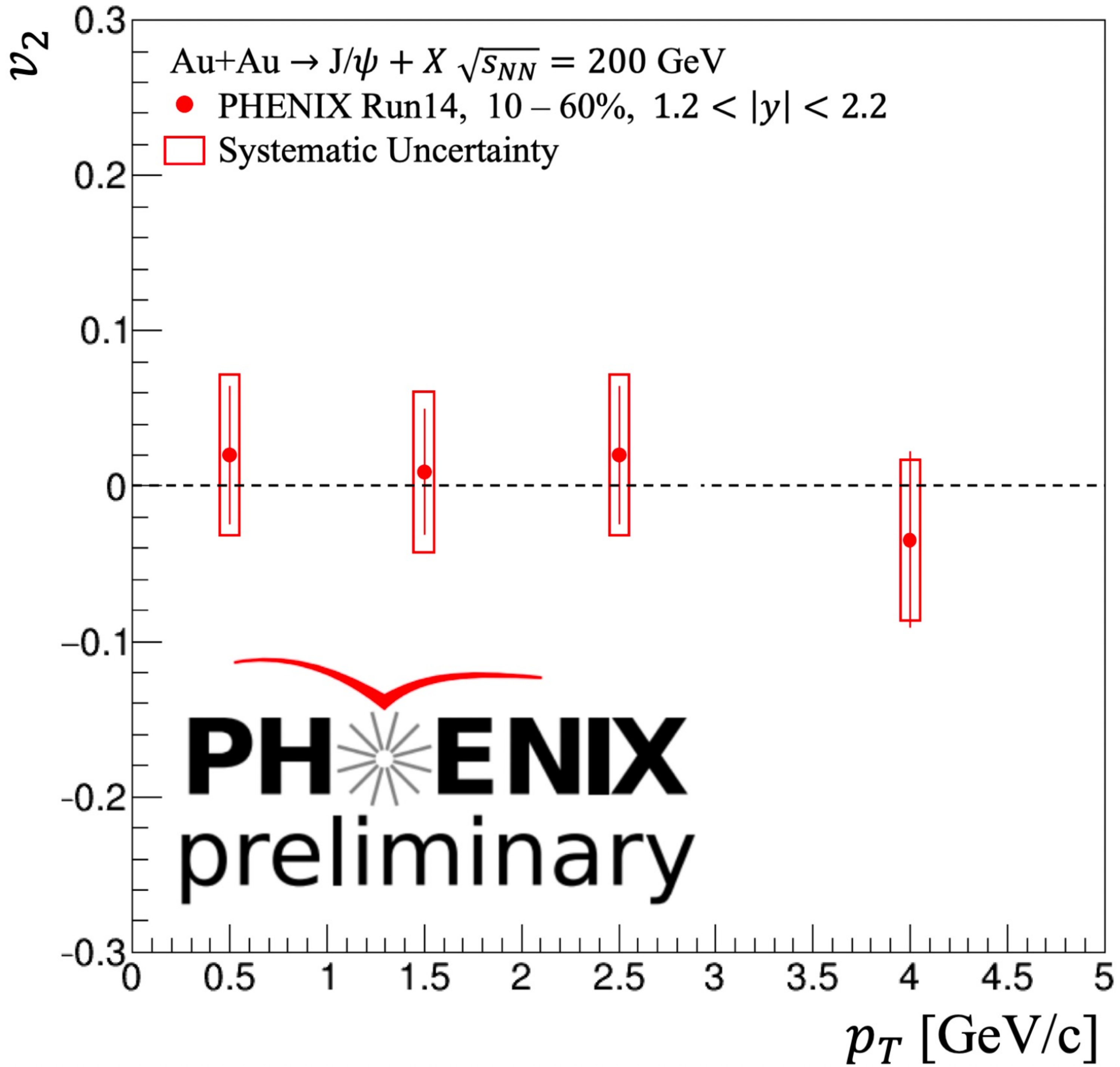}}
\caption{Elliptic flow of $J/\psi$ as a function of $p_T$ at forward rapidity in 10--60\% Au+Au collisions at $\sqrt{s_{NN}}=200$ GeV.}
\label{Fig:large_jpsi}
\end{figure}

In early studies of $J/\psi$ production and its modification in heavy-ion collisions at RHIC and LHC energies, it was found that the regeneration effect is important at the LHC energy but may not at the RHIC energy~\cite{PHENIX:2011img,ALICE:2012jsl}.
In addition to the nuclear modification factor, elliptic flow ($v_2$) can provide more information on mechanisms of the $J/\psi$ production in heavy-ion collisions.
PHENIX has measured $J/\psi$ $v_2$ at forward rapidity in 10--60\% Au+Au collisions at $\sqrt{s_{NN}}=200$ GeV as presented in Fig.~\ref{Fig:large_jpsi}, and it is consistent with zero within uncertainties.
The PHENIX results are different from the non-zero $v_2$ results at the LHC energies~\cite{ALICE:2018bdo}, and this indicates that the regeneration effect is small at forward rapidity at the RHIC energy.

\section{Summary and outlook}
In summary, PHENIX has obtained results indicating QGP droplets in small systems such as collectivity, nuclear modification of light-flavor hadrons, and $\psi(2S)$ in high multiplicity events.
In large systems, detailed studies of the QGP have been done with direct photons, particle correlations, heavy-flavor electrons, and quarkonia.
PHENIX recently completed the data production of all collected data, so the full statistics of Au+Au data collected in 2014 and 2016 are ready for analysis in the future.


\begin{thebibliography}{0}%
\makeatletter
\providecommand \@ifxundefined [1]{%
 \@ifx{#1\undefined}
}%
\providecommand \@ifnum [1]{%
 \ifnum #1\expandafter \@firstoftwo
 \else \expandafter \@secondoftwo
 \fi
}%
\providecommand \@ifx [1]{%
 \ifx #1\expandafter \@firstoftwo
 \else \expandafter \@secondoftwo
 \fi
}%
\providecommand \natexlab [1]{#1}%
\providecommand \enquote  [1]{``#1''}%
\providecommand \bibnamefont  [1]{#1}%
\providecommand \bibfnamefont [1]{#1}%
\providecommand \citenamefont [1]{#1}%
\providecommand \href@noop [0]{\@secondoftwo}%
\providecommand \href [0]{\begingroup \@sanitize@url \@href}%
\providecommand \@href[1]{\@@startlink{#1}\@@href}%
\providecommand \@@href[1]{\endgroup#1\@@endlink}%
\providecommand \@sanitize@url [0]{\catcode `\\12\catcode `\$12\catcode
  `\&12\catcode `\#12\catcode `\^12\catcode `\_12\catcode `\%12\relax}%
\providecommand \@@startlink[1]{}%
\providecommand \@@endlink[0]{}%
\providecommand \url  [0]{\begingroup\@sanitize@url \@url }%
\providecommand \@url [1]{\endgroup\@href {#1}{\urlprefix }}%
\providecommand \urlprefix  [0]{URL }%
\providecommand \Eprint [0]{\href }%
\providecommand \doibase [0]{https://doi.org/}%
\providecommand \selectlanguage [0]{\@gobble}%
\providecommand \bibinfo  [0]{\@secondoftwo}%
\providecommand \bibfield  [0]{\@secondoftwo}%
\providecommand \translation [1]{[#1]}%
\providecommand \BibitemOpen [0]{}%
\providecommand \bibitemStop [0]{}%
\providecommand \bibitemNoStop [0]{.\EOS\space}%
\providecommand \EOS [0]{\spacefactor3000\relax}%
\providecommand \BibitemShut  [1]{\csname bibitem#1\endcsname}%
\let\auto@bib@innerbib\@empty
%</preamble>
\end{thebibliography}%


\begin{thebibliography}{99}
\bibitem{PHENIX:2018lia}
C. Aidala, {\it et al.}  
(PHENIX Collaboration),
Creation of quark gluon plasma droplets with three distinct geometries
Nature Phys. {\bf 15}, 214 (2019).

\bibitem{PHENIX:2021ubk}
U. Acharya, {\it et al.} 
(PHENIX Collaboration),
Kinematic dependence of azimuthal anisotropies in 
$p+Au,$ $d+Au,$ and $^3He+Au$ at $\sqrt{s}_{NN}=200$ GeV,
Phys. Rev. C {\bf 105}, 024901 (2022).

\bibitem{PHENIX:2022nht}
U. Acharya, {\it et al.} 
(PHENIX Collaboration),
Measurements of second-harmonic Fourier coefficients from 
azimuthal anisotropies in $p$$+$$p$, $p$+Au, $d$+Au, and $^3$He+Au 
collisions at $\sqrt{s_{_{NN}}}=200$ GeV,
arXiv:2203.09894.

\bibitem{PHENIX:2021dod}
U. Acharya, {\it et al.} 
(PHENIX Collaboration),
Systematic study of nuclear effects in 
$p$$+$Al, $p$$+$Au, $d$$+$Au, and $^{3}$He$+$Au collisions at
$\sqrt{s_{_{NN}}}=200$ GeV using $\pi^0$ production,
Phys. Rev. C {\bf 105}, 064902 (2022).

\bibitem{PHENIX:2022hku}
U. Acharya, {\it et al.} 
(PHENIX Collaboration),
Study of $\phi$-meson production in $p$$+$Al, $p$$+$Au, $d$$+$Au, and $^3$He$+$Au 
collisions at $\sqrt{s_{_{NN}}}=200$ GeV.
Phys. Rev. C {\bf 106}, 014908 (2022).

\bibitem{PHENIX:2019brm}
U. Acharya, {\it et al.} 
(PHENIX Collaboration),
Measurement of $J/\psi$ at forward and backward rapidity in
$p$$+$$p$, $p$$+$Al, $p$$+$Au, and $^3$He$+$Au collisions at
$\sqrt{s_{_{NN}}}=200$ GeV,
Phys. Rev. C {\bf 102}, 014902 (2020).

%nuclear modification of psi(2S) 
\bibitem{PHENIX:2022nrm}
U. Acharya, {\it et al.} 
(PHENIX Collaboration),
Measurement of $\psi(2S)$ nuclear modification at backward and
forward rapidity in $p$$+$$p$, $p$$+$Al, and $p$$+$Au collisions at
$\sqrt{s_{_{NN}}}=200$ GeV
Phys. Rev. C {\bf 105}, 064912 (2022).

\bibitem{PHENIX:2022qfp}
U. Acharya, {\it et al.} 
(PHENIX Collaboration),
Low-$p_T$ direct-photon production in Au$+$Au collisions at
$\sqrt{s_{_{NN}}}=39$ and 62.4 GeV,
arXiv:2203.12354.

%non-prompt direct photon
\bibitem{PHENIX:2022rsx}
U. Acharya, {\it et al.} 
(PHENIX Collaboration),
Nonprompt direct-photon production in Au$+$Au collisions at
$\sqrt{s_{_{NN}}}=200$ GeV,
arXiv:2203.17187.

\bibitem{ALICE:2015xmh}
J. Adam, {\it et al.} 
(ALICE Collaboration),
Direct photon production in Pb-Pb collisions at
$\sqrt{s_{NN}}=$ 2.76 TeV,
Phys. Lett. B {\bf 754}, 235 (2016).

%heavy-flavor electron in pp
\bibitem{PHENIX:2019pxh}
U. Acharya, {\it et al.} 
(PHENIX Collaboration),
Measurement of charm and bottom production from semileptonic
hadron decays in $p$$+$$p$ collisions at $\sqrt{s_{NN}}=200$ GeV,
Phys. Rev. D {\bf 99}, 092003 (2019).

%heavy-flavor electron in AuAu
\bibitem{PHENIX:2022wim}
U. Acharya, {\it et al.} 
(PHENIX Collaboration),
Charm- and Bottom-Quark Production in Au$+$Au Collisions at
$\sqrt{s_{_{NN}}}$ = 200 GeV,
arXiv:2203.17058.

\bibitem{PHENIX:2011img}
A. Adare, {\it et al.} 
(PHENIX Collaboration),
$J/\psi$ suppression at forward rapidity in Au$+$Au collisions
at $\sqrt{s_{NN}}=200$ GeV,
Phys. Rev. C {\bf 84}, 054912 (2011).

\bibitem{ALICE:2012jsl}
B. Abelev, {\it et al.} 
(ALICE Collaboration),
$J/\psi$ suppression at forward rapidity in Pb-Pb collisions
at $\sqrt{s_{NN}}=2.76$ TeV,
Phys. Rev. Lett. {\bf 109} 072301 (2012).

\bibitem{ALICE:2018bdo}
S. Acharya, {\it et al.} 
(ALICE Collaboration),
Study of J/$\psi$ azimuthal anisotropy at forward rapidity in
Pb-Pb collisions at $\sqrt{s_{_{NN}}}=5.02$ TeV,
J. High Energy Phys. {\bf 02 (2019)} 012.

\end{thebibliography}
\end{document}